\documentstyle[12pt]{article}

\advance \topmargin by -\headheight
\advance \topmargin by -\headsep
     
\evensidemargin \oddsidemargin
\def\rf#1{(\ref{eq:#1})}
\def\lab#1{\label{eq:#1}}
\def\nonu{\nonumber}
\def\br{\begin{eqnarray}}
\def\er{\end{eqnarray}}
\def\be{\begin{equation}}
\def\ee{\end{equation}}

\def\lb{\lbrack}
\def\rb{\rbrack}

\def\llb{\left\lbrack}
\def\rrb{\right\rbrack}
\def\Blb{\Bigl\lbrack}
\def\Brb{\Bigr\rbrack}
\def\lcurl{\left\{}
\def\rcurl{\right\}}
\def\({\left(}
\def\){\right)}
\def\v{\vert}                     
\def\bgv{\bigg\vert}              %%
\def\lskip{\vskip\baselineskip\vskip-\parskip\noindent}
\def\mskp{\par\vskip 0.3cm \par\noindent}

\def\bc{\begin{center}}
\def\ec{\end{center}}
\newcommand{\sect}[1]{\setcounter{equation}{0}\section{#1}}

\relax
\newcommand\partder[2]{{{\partial {#1}}\over{\partial {#2}}}}

\newcommand\partderh[3]{{{\partial^{#3} {#1}}\over{{\partial {#2}}^{#3} }}}

\newcommand\me[2]{\left\langle {#1}\right|\left. {#2} \right\rangle} 
\newcommand\sbr[2]{\left\lbrack\,{#1}\, ,\,{#2}\,\right\rbrack} 
\newcommand\Sbr[2]{\Bigl\lbrack\,{#1}\, ,\,{#2}\,\Bigr\rbrack}

\def\a{\alpha}

\def\c{\chi}
\def\d{\delta}
\def\D{\Delta}

\def\l{\lambda}

\def\o{\over}

\def\p{\phi}
\def\P{\Phi}
\def\bp{{\bar \p}}
\def\pa{\partial}
\def\pr{\prime}

\def\t{\tau}

\def\ti{\tilde}
\def\tit{{\tilde t}}
\def\wti{\widetilde}
\newcommand\sumi[1]{\sum_{#1}^{\infty}}

\def\cG{{\cal G}}
\def\cL{{\cal L}}

\def\cW{{\cal W}}
\def\cS{{\cal S}}

\font \msb=msbm10 scaled \magstep1

\newcommand{\IC}{\mbox{\msb C} }

\def\one{\hbox{{1}\kern-.25em\hbox{l}}}
\def\mark{\noindent{\bf Remark.}\quad}

\newtheorem{definition}{Definition}[section]

\newtheorem{lemma}{Lemma}[section]
\newtheorem{corollary}{Corollary}[section]
\def\proof{\par{\it Proof}. \ignorespaces} \def\endproof{{$\Box$}\par}
 
\def\cKP{{\sf cKP}~}
\def\cKPrm{${\sf cKP}_{r,m}$~}

\newcommand\DB{{Darboux-B\"{a}cklund}~}

\newcommand{\nit}{\noindent}
\newcommand{\ct}[1]{\cite{#1}}
\newcommand{\bi}[1]{\bibitem{#1}}

\newcommand\CMP[3]{{\sl Commun. Math. Phys.} {\bf #1} (#2) #3}

\newcommand\JMP[3]{{\sl J. Math. Phys.} {\bf #1} (#2) #3}

\newcommand\FAaIA[3]{{\sl Functional Analysis and Its Application} {\bf #1}
(#2) #3}
\newcommand\LMP[3]{{\sl Letters in Math. Phys.} {\bf #1} (#2) #3}
\newcommand\IJMPA[3]{{\sl Int. J. Mod. Phys.} {\bf A#1} (#2) #3}

\newcommand\PHSA[3]{{\sl Physica} {\bf A#1} (#2) #3}

\begin{document}

\vspace*{-.5cm}
\noindent
{\sl solv-int/9805006} \\%\hfill{UICHEP-TH/98-xx}\\

\begin{center}
{\Large {\bf On Grassmannian Description of the Constrained KP Hierarchy}}
\end{center}
\vskip .15in
\begin{center}
{ H. Aratyn}
\par \vskip .1in \noindent
Department of Physics, University of Illinois at Chicago\\
845 W. Taylor St., Chicago, IL 60607-7059
\end{center}

\begin{abstract}
This note develops an explicit construction of the constrained KP hierarchy 
within the Sato Grassmannian framework.
Useful relations are established between the kernel elements of the underlying 
ordinary differential operator and the eigenfunctions of the associated KP 
hierarchy as well as between the related bilinear concomitant 
and the squared eigenfunction potential. 
\end{abstract}

{\sect{Introduction}}    
    
The purpose of this note is to present construction of the constrained KP
(\cKP) hierarchy within the Sato Grassmannian context
using elements of the kernels of the underlying differential operators.
The fundamental concept is the canonical pairing (the bilinear concomitant) 
introduced here on the space
of elements of the kernels of the underlying differential operator and
its conjugated counterpart.
The formalism is simplified by relations between the bilinear concomitant
and the squared eigenfunction potential (SEP) which emerged before in 
the setting of the KP hierarchy \ct{oevel93,sep}.
The claim is that use of SEP makes construction of the \cKP hierarchy 
within the Sato Grassmannian theory of the KP hierarchy more transparent.
The \cKP hierarchy has recently been discussed in \ct{Z} and \ct{HL,HLa}
using Segal-Wilson modification of the Sato Grassmannian. 
This note provides the link between these works and the current formalism 
based on the Sato Grassmannian and the SEP method.

{\sect{KP Hierarchy}}

We first briefly review the KP hierarchy of nonlinear evolution equations
in the approach based on the calculus of the pseudodifferential operators.
The main object here is the pseudo-differential Lax operator $L$ :
\be
L = D^r + \sum_{j=0}^{r-2} v_j D^j + \sum_{i \geq 1} u_i D^{-i}
\lab{gen-KP}
\ee
The operator $D$ satisfies the generalized Leibniz rule so for instance
$ \sbr{D}{f} = f^{\pr}$ with $f^{\pr}= \pa f = \pa f /\pa x$.

The associated isospectral flows are described by the Lax equations :
\be
\partder{}{t_n} L = \Sbr{L^{n / r}_{+}}{L}
\quad    \; \; n = 1, 2, \ldots
\lab{lax-eq}
\ee
with $x \equiv t_1$.
In \rf{lax-eq} and below, the subscripts $(\pm )$ of pseudo-differential
operators indicate projections on purely 
differential/pseudo-differential parts.
Commutativity of the isospectral flows $\pa/ \pa {t_n}$ \rf{lax-eq}
is then assured by the Zakharov-Shabat equations.

For a given Lax  operator $L$, which satisfies Sato's flow equation
\rf{lax-eq}, we call the function $\Phi$ ($ \Psi$), whose flows are
given by the expression:
\be
\partder{\Phi}{t_l} = L^{l \o r}_{+} (\Phi ) \qquad; \qquad
\partder{\Psi}{t_l} = - \( L^{\ast} \)^{l \o r}_{+} (\Psi )
\quad\;\; l=1,2, \ldots
\lab{eigenlax}
\ee
an {\it (adjoint) eigenfunction} of $L$. 
In \rf{eigenlax} we have
introduced an operation of conjugation, defined by simple rules
$D^{\ast} = -D$ and $(AB)^{\ast}= B^{\ast} A^{\ast}$.
Throughout this paper we will follow the convention that for any 
(pseudo-)differential operator $A$ and a function $f$, the symbol $A(f)$ 
will indicate application (action) of $A$ on $f$ while the symbol 
$Af$ will be just a product of $A$ with the zero-order (multiplication) 
operator $f$.

One can also represent the Lax operator in terms of the dressing
operator $W= 1 + \sum_1^{\infty} w_n D^{-n}$ through
$L= W \, D^r \,W^{-1}$.
In this framework equation \rf{lax-eq} is equivalent to the so called
Wilson-Sato equation:
\be
\pa_n  W = - \( W D^n W^{-1} \)_{-} W
\lab{sato-a}
\ee
where $\pa_n = \pa / \pa t_n$.
Next, we define  corresponding wave-eigenfunction via:
\be
\psi_W (t,\l ) = W \bigl( e^{\xi (t,\l )}\bigr) = 
\(1 + \sum_{i=1}^{\infty} w_i(t)\l^{-i} \)\, e^{\xi (t,\l )}
\lab{BA}
\ee
where
\be
\xi(t,\lambda) \equiv  \sum_{n=1}^\infty t_n\lambda^n \qquad; \quad
t_1 = x
\lab{xidef}
\ee
Similarly, there is also an adjoint wave-eigenfunction:
\be
\psi^{*}_W  = W^{*-1}\bigl( e^{-\xi (t,\l )}\bigr) = 
\( 1 + \sum_{i=1}^{\infty} w_i^{*}(t)\lambda^{-i}\) e^{-\xi (t,\l )}     
\lab{BA-adjoint} 
\ee
As seen from \rf{sato-a} and \rf{BA} the wave-eigenfunction, is an
eigenfunction which in addition to eqs.\rf{eigenlax} also satisfies
the spectral equations $L \psi_W (\l ,t) = \l^r \psi_W (\l ,t)$.
The wave-eigenfunction and its adjoint enter the fundamental Hirota's 
bilinear identity:
\be
\int  d\l \,\psi_{W}^{*}(t,\l)  \psi_{W}(t^{\pr} ,\l) = 0
\lab{bilide}
\ee
which generates the entire KP hierarchy via Hirota's equations for the
underlying tau-functions (see f.i. \ct{cortona}).
In \rf{bilide} and in what follows integrals over spectral parameters
are understood as:
$\int d\l \equiv \oint_{0} \frac{d\l}{2i\pi} ={\rm Res}_{\l = 0}$.
The proper understanding of \rf{bilide} requires, following f.i. 
\ct{cortona}, expanding of $\psi_{W}(t^{\pr} ,\l)$
in \rf{bilide}
as formal power series w.r.t. $t^{\pr}_n -t_n$, $n=1,2,\ldots $
according to 
\be
\psi_{W} (t^{\pr}) = \sum {\( t_1^{\pr} - t_1\)^{k_1} {\cdots} 
\( t_n^{\pr} - t_n\)^{k_n} \o  {k_1}! {\cdots} {k_n}!}
\pa_1^{k_1} {\cdots} \pa_n^{k_n} \psi_{W} (t)
\lab{expand}
\ee

The wave function is an oscillatory function of order $0$.
Generally, the oscillatory function of order $l$ is of the form:
\be
f( t, \l) e^{\xi (t,\l )} = \( \l^l + \sum_{j<l} a_j (t) \l^j \)
 \, e^{\xi (t,\l )}
\lab{def-osci}
\ee
It will be of importance for us that the action of the differential operator
$D$ can be uniquely inverted on the space of oscillatory functions according 
to
\be
D^{-1} f( t, \l) e^{\xi (t,\l )} = \sumi{\a=0} (-1)^{\a} f^{(\a)} ( t, \l)
\l^{-1-\a} e^{\xi (t,\l )}
\lab{dinverse}
\ee

Consider now the one-form:
$ \omega = \sum_n {\rm Res} \( D^{-1} \Psi L^{n/r}_{+} \P  D^{-1} \) dt_n$
defined for the couple of (adjoint) eigenfunctions $\Phi, \Psi$.
One shows \ct{oevel93} using the Zakharov-Shabat equations that $\omega $
is a closed form with respect to the exterior derivative
$d \equiv \sum_n \pa_n d t_n$.
By the usual argument one concludes from $d \omega =0$ that the one form
$\omega$ can be rewritten as $ \omega = d S (\P, \Psi)$.
This procedure defines (up to a constant) a squared
eigenfunction potential (SEP) $S (\P, \Psi)$. 
In particular, the flows of $S (\P, \Psi)$ are given by
\be
\pa_n S (\P, \Psi)= {\rm Res} \( D^{-1} \Psi L^{n/r}_{+} \P  D^{-1} \)
\lab{sepflows}
\ee
Especially, 
$\pa_x S\( \P (t), \Psi (t) \) = \P (t) \Psi (t)$.
As shown in \ct{sep} the squared
eigenfunction potential defines a spectral
representation of (adjoint) eigenfunctions.
The statement is as follows.
Any (adjoint) eigenfunction of the general KP hierarchy possesses a spectral
representation:
\br
\P (t) &=& - \int d\l\,\psi_{W} (t ,\l )\, 
S\(\P (t^\pr ), \psi_{W}^{\ast} (t^\pr ,\l )\)
\lab{spec1}\\
\Psi (t) &=& \int d \l\,\psi_{W}^{\ast} (t ,\l )\, 
S\(\psi_{W} (t^\pr ,\l ), \Psi (t^\pr )\)
\lab{spec2}
\er
with spectral densities given by the squared eigenfunction potentials
at some multi-time $t^{\pr} = \( t_1^{\pr},t_2^{\pr},\ldots \)$
taken at some arbitrary fixed value. 
The r.h.s. of \rf{spec1} and \rf{spec2} do not depend on $t^{\pr}$.
Furthermore, the closed expressions have been found in \ct{sep}
for those squared eigenfunction potentials which have as argument at least one 
oscillating wave-eigenfunction:
\br
S \(\P (t),\psi^{\ast}_{W} \(t\)\) &=& - { 1\o \l} \psi^{\ast}_{W} \(t\)
\P \( t + [\l^{-1}]\)
\lab{whpd} \\
S \(\psi_{W} \(t\) ,\Psi (t)\)  &=&{ 1\o \l} \psi_{W} \(t\) 
\Psi \(t - [\l^{-1}]\)
\lab{whpe}
\er
In the above equation $S \( \psi_W (t, \l), \Psi (t)\)$
is the squared eigenfunction potential (SEP) associated
with a pair of eigenfunctions $\psi_W (t, \l)$ and $\Psi (t)$.
It is an oscillatory function of order $-1$:
\be
S\(\psi_W (t, \l), \Psi (t) \) = 
\sum_{j=1}^\infty s_{j} (t) \l^{-j} \, e^{\xi (t,\l )}= 
 \llb \Psi (t) \l^{-1} + O(\l^{-2}) \rrb \, e^{\xi (t,\l )}
\lab{S-osci}
\ee

We will now make connection to the language of universal Sato
Grassmannian ${\cG}r$. 
Consider the hyperplane ${\bf W} \in {\cG}r$
defined through a linear basis of Laurent series
$\lcurl f_k (\l ) \rcurl$ in $\l$ in terms of the wave eigenfunction 
as a generating function :
\br
{\bf W} &\equiv& {\rm span } \langle f_0 (\l )\, ,\,f_1 (\l )\, ,\, 
f_2 (\l )\,, \ldots \rangle \nonu \\
f_k (\l ) &=& \partderh{}{x}{k} \psi_W (t,\l ) \bgv_{x= t_2 =t_3 = \ldots =0}
\lab{wgras}
\er
Obviously, ${\bf W}$ is closed under the differentiation $\pa / \pa x$.
{}From the fact that $\psi_W (t, \l)$ satisfies eq.\rf{eigenlax}
we obtain an alternative definition of ${\bf W}$:
\be
{\bf W} = {\rm span} \{ \psi_W (t, \l),~ {\rm all}\;\; t \in \IC^{\infty} \}
\lab{wgras-alt}
\ee
A typical element of ${\bf W}$; 
$f_k (\l )= \( \l^k + O (\l^{k-1}) \) \exp \xi(t,\lambda) $,
has an order $k \geq 0$.
Consequently, the set of orders of all of elements of ${\bf W}$ is given by 
the set of non-negative integers.

In case of the standard $r$-th KdV reduction, where the corresponding
Lax operator $\cL = D + \sumi{1} u_i D^{-i}$ satisfies $\cL^r = \cL^r_{+}$,
the latter constraint translates to the Grassmannian language as
$\l^r {\bf W} \subset {\bf W}\,$. 

It is clear that $ \int  d\l \,\psi_{W}^{*}(t,\l)  \c (t^{\pr} ,\l) =0$
for any $\c (t, \l) \in {\bf W}$.
We will make here a plausible assumption that the inverse holds as well.
More precisely, the statement is as follows. Let
$F \(\psi_W (t^{\pr} ,\l)\)$ be a linear functional of
$\psi_W (t ,\l)$ of a positive order for which the following bilinear equation
$ \int  d\l \,\psi_{W}^{*}(t,\l)  F \(\psi_W (t^{\pr} ,\l)\) = 0$  holds
for all $t^{\pr}$, then  $F \(\psi_W (t^{\pr} ,\l)\)\in {\bf W}$.

{\sect{Differential Operators and the Canonical Pairing Str\-uct\-ure}}

Consider a differential operator of order $m$
\be
L_m = D^m +u_{m-1} D^{m-1}+ \ldots + u_{1} D^{1} +u_0
\lab{lm-def}
\ee
The differential operator of order $m$ is called
{\em monic} if its leading term is $D^m$.
A monic differential operator of order $m$ is fully characterized by
 $m$ elements of its kernel.
For instance, let functions $\p_i\;,\; i=1,{\ldots} ,m$ constitute
a basis for ${\rm Ker}\, L_m =\{\p_1, \ldots , \p_m\}$ then 
\be
L_m (f) =  {W_{m} \lb \p_1, \ldots , \p_{m}, f \rb 
  \o W_{m} \lb \p_1, \ldots , \p_{m}\rb}
\lab{lmdef}
\ee
The elements of ${\rm Ker}\, L_m$ are assumed to be 
linearly independent so that the Wronskian matrix:
\be
\( \cW_{m \times m} \)_{1 \leq i,j\leq m} = \pa_x^{j-1} \p_i
\lab{wrdef}
\ee
is invertible.
In different words the Wronskian determinant 
$W_{m} \lb \p_1, \ldots , \p_{m} \rb = \det \v\v \cW_{m \times m} \v \v $
must be different from zero.
We define 
$ \(\cW^{-1}_{m \times m}\)_{ij} \; i,j=0, {\ldots} , m-1  $ to be 
the matrix elements of the matrix of the inverse of the Wronskian matrix
$\cW_{m \times m}$. The following relations are then satisfied:
\be
\sum_{j=1}^{m} \(\cW^{-1}_{m \times m}\)_{ij} \p^{(j-1)}_k = \d_{i,k} \quad ; 
\quad \sum_{k=1}^{m} \p^{(j-1)}_k \(\cW^{-1}_{m \times m}\)_{kl} = \d_{j,l}
\lab{deltas}
\ee
It is easy to verify that 
\be
\(\cW^{-1}_{m \times m}\)_{ij} = (-1)^{i+j} { \det_{(j,i)} \v\v 
\cW_{m \times m} \v\v \o W_{m} \lb \p_1, \ldots , \p_{m} \rb }
\lab{winverse}
\ee
where the determinant on the right hand side is the minor determinant 
obtained by extracting the $j$'th row and $i$'th column
from the Wronski matrix $\cW_{m \times m}$ given in eq.\rf{wrdef}.

The following technical identity, which is valid for an arbitrary function
$\c$, follows directly from \rf{wrdef}-\rf{winverse}:
\be
\sum_{j=1}^{m} \(\cW^{-1}_{m \times m}\)_{ij} \c^{(j-1)}=
(-1)^{m+i} { W_{m} \lb \p_1, \ldots , {\widehat \p_{i}}, \ldots 
\p_m, \c \rb \o W_{m} \lb \p_1, \ldots , \p_{m}\rb } \quad ;\quad
i=1,\ldots ,m
\lab{winvc}
\ee

In addition, we also need to consider an adjoint operator $L_m^{\ast}$
obtained from \rf{lm-def} by a process of conjugation described below 
eq.\rf{eigenlax}.
Let $\psi_1, \ldots , \psi_m$ be elements of the kernel of an 
adjoint operator $L_m^{\ast}$:
\be
{\rm Ker}\, L_m^{\ast}= \{\psi_1, \ldots , \psi_m \} 
\lab{lmadef}
\ee
They are given in terms of elements of ${\rm Ker}\, L_m $ as follows 
\ct{oevel96,v502}: 
\be
\psi_i = (-1)^{m+i} { W_{m-1} \lb \p_1, \ldots , {\widehat \p_{i}}, \ldots 
\p_m  \rb \o W_{m} \lb \p_1, \ldots , \p_{m}\rb } \quad ;\quad
\, i=1,\ldots ,m
\lab{psiipi}
\ee
Comparing with \rf{winverse} we see that the relation \rf{psiipi} expresses 
the fact that $(\psi_1, \ldots ,\psi_m )^T$ is the last column 
in the inverse $\cW^{-1}$ of the Wronskian matrix $\cW$ 
of $(\p_1, \ldots ,\p_m )$.
In particular, we see that the functions 
$\{\psi_1, \ldots ,\psi_m \}$ are also linearly independent.

Some of the obvious consequences of definition \rf{psiipi} and connection
between $\psi_i$ and the matrix $\cW^{-1}$ are:
\be
\sum_{i=1}^m \p_i^{(k)} (t) \psi_i (t) = \d_{k,m-1} \quad{\rm for} \quad
k=0,1,\ldots ,m-1
\lab{wrona}
\ee
For completeness let us list the extension of \rf{wrona} to $k=m$:
\br
\sum_{i=1}^m \p_i^{(m)} (t) \psi_i (t) & = &
\sum_{i=1}^m (-1)^{m-i} { W_{m-1} \lb \p_1, \ldots , {\widehat \p_{i}}, \ldots 
\p_m  \rb \p_i^{(m)} \o W_{m} \lb \p_1, \ldots , \p_{m}\rb } \nonu \\
&= &\pa_x \ln W_{m} \lb \p_1, \ldots , \p_{m}\rb
\lab{wronb}
\er
Consider the quantity $N =\sum_{i=1}^m \, \p_i\, D^{-1} \, \psi_i$.
It follows easily that
\be
( L_m N)_{-} = \sum_{i=1}^m \, L_m (\p_i)\, D^{-1} \, \psi_i=0
\lab{nmin}
\ee
Moreover, using the Leibniz rule we obtain from \rf{wrona} and \rf{wronb}
\br
( L_m N)_{+} &=& \( L_m (\sum_{i=1}^m \sum_{\a=1}^{\infty} D^{-1-\a} 
\p_i^{(\a)}\,  \psi_i) \)_{+} \lab{wronc}\\
&=& \( L_m ( D^{-m} + D^{-1-m} \pa_x \ln W_{m} \lb \p_1, \ldots , \p_{m}\rb
+ O (D^{-2-m}) ) \)_{+} = 1
\nonu
\er
Hence, as in \ct{oevel96} we obtain from \rf{nmin} and \rf{wronc} 
\be
L_m^{-1} \; = \; \sum_{i=1}^m \, \p_i\, D^{-1} \, \psi_i
\lab{lminv}
\ee
Consider, now
${\rm Res} \( D^{-1} \psi_j L_m \sum_{i=1}^m \, \p_i\, D^{-1} \, \psi_i \)$.
In view of \rf{lminv} we find:
\be
{\rm Res} \( D^{-1} \psi_j L_m \sum_{i=1}^m \, \p_i\, D^{-1} \, \psi_i \)
= \psi_j =
\sum_{i=1}^m {\rm Res} \( D^{-1} \psi_j L_m \, \p_i\, D^{-1} \, \) \psi_i
\lab{resapsi}
\ee
One notices that 
\be
\pa_x {\rm Res} \( D^{-1} \psi_j L_m \, \p_i\, D^{-1} \, \) 
= \psi_j L_m (\p_i) + L_m^{\ast} (\psi_j ) \, \p_i =0
\lab{consta}
\ee
and therefore ${\rm Res} \( D^{-1} \psi_j L_m \, \p_i\, D^{-1} \, \)$ is a
constant in $x$.
Since, functions $\psi_i$ are linearly independent, we conclude in view of 
equation \rf{resapsi} and \rf{consta} that:
\be
{\rm Res} \( D^{-1} \psi_j \, L_m \, \p_i\, D^{-1} \, \) = \d_{i,j}
\lab{resdij}
\ee
It appears, therefore, that $\{\psi_1, \ldots ,\psi_m \}$ can be viewed 
as the dual basis of $\{\p_1, \ldots ,\p_m \}$ with respect to a 
canonical pairing defined in terms of the so-called bilinear concomitant 
(see \ct{ince,wilson-crm,wilson-agd}):
\br
{\me{\p}{\psi}}_{L_m} & \equiv &
{\rm Res} \( D^{-1} \psi L_m \p D^{-1} \) \lab{bilres}\\
&=&
\sum_{i=1}^m \sum_{j=0}^{i-1}\, (-1)^j \, \p^{(i-j-1)}\, \( u_i \psi\)^{(j)} 
\lab{bil}
\er
with $u_m=1$.
In this setting the bases $\{\p_1, \ldots ,\p_m \}$ and
$\{\psi_1, \ldots ,\psi_m \}$ related through \rf{psiipi} are
dual to each other in the sense of satisfying
${\me{\p_i}{\psi_j}}_{L_m} = \d_{ij}$  for $i,j=1,\ldots , m$
due to \rf{resdij}.
The following technical Lemma provides a useful characterization of the
products ${\me{\c}{\psi_i}}_{L_m}$ for an arbitrary function $\c$
and $\psi_i \in {\rm Ker} L_m^{\ast}$.
\begin{lemma}
The following identity:
\be
{\me{\c}{\psi_i}}_{L_m} = 
\sum_{j=1}^{m} \(\cW^{-1}_{m \times m}\)_{ij} \c^{(j-1)}
\quad ;\quad
i=1,\ldots ,m
\lab{psiiclm}
\ee
holds for an arbitrary function $\c$
and $\psi_i \in {\rm Ker} L_m^{\ast}$
\label{lemma:psiiclm}
\end{lemma}
\begin{proof}
Let $M_i = \sum_{j=1}^{m} \(\cW^{-1}_{m \times m}\)_{ij} D^j$ be $(m-1)$-order
differential operator for the fixed $i$.
We know its $m-1$ null-functions $\p_k$ such that
$M_i ( \p_k)=0$ for $k\ne i$. We also have $M_i (\p_i)=1$.
This characterizes $M_i$ completely.
Note, that the $(m-1)$-order differential operator
$ {\me{\cdot}{\psi_i}}_{L_m}$ agrees with $M_i$ on 
$\p_i\, , \, i=1, \ldots , m$, which completes the proof.
\end{proof}
Recalling identity \rf{winvc} we find an alternative way of writing 
\rf{psiiclm} as
\be
{\me{\c}{\psi_i}}_{L_m} = 
(-1)^{m+i} { W_{m} \lb \p_1, \ldots , {\widehat \p_{i}}, \ldots 
\p_m, \c \rb \o W_{m} \lb \p_1, \ldots , \p_{m}\rb } 
\lab{inversebi}
\ee
from which it follows:
\begin{corollary}
\be 
{\me{\c}{\psi_i}}_{L_m} =  \psi_i L_{m,i} \( \c \)
\lab{psilmic}
\ee
where $L_{m,i} $ are the ordinary differential operators of order $m-1$,
whose kernels are given by
$ {\rm Ker} \( L_{m,i}  \) = 
\{ \p_1, \ldots , {\widehat \p_{i}}, \ldots, \p_m \}$.
Correspondingly, the action of $L_{m,i} $ is defined through 
\be
L_{m,i}  \( \c \) 
\equiv { W_{m+1} \lb \p_1, \ldots , {\widehat \p_{i}}, \ldots, \p_m , \c \rb 
  \o W_{m} \lb \p_1, \ldots , {\widehat \p_{i}}, \ldots , \p_{m}\rb } 
\lab{lmic}
\ee
\label{corollary:lmic}
\end{corollary}

We will now introduce isospectral deformations of the
differential operator $L_m$ of the form:
\be
\pa_n L_m = {\wti B}_n L_m - L_m B_n \quad ;\quad n=1,2, \ldots 
\lab{lmflows}
\ee
In this setting we will show that the product ${\me{\cdot}{\cdot}}_{L_m} $ 
defines a canonical pairing 
$ {\rm Ker} \,  L_m \times {\rm Ker} \,L_m^{\ast} \to \IC$.
As discussed in \ct{wilson-crm,wilson-agd} this pairing  is nonsingular.

The two families of differential operators ${\wti B}_n, B_n$ are both 
assumed to satisfy Zakharov-Shabat equations:
\br
0 &=& \pa_k {\wti B}_n - \pa_n {\wti B}_k + \sbr{{\wti B}_n}{{\wti B}_k} 
\nonu\\
0 &=& \pa_k  { B}_n - \pa_n { B}_k + \sbr{{ B}_n}{{ B}_k} 
\quad ;\quad k,n=1,2, \ldots 
\lab{zsflows}
\er
to ensure commutativity of flows defined in \rf{lmflows}.
{}From \rf{lmflows} we find  that 
\be
\pa_n L_m^{-1} = B_n L_m^{-1} - L_m^{-1} {\wti B}_n \quad ;\quad n=1,2, \ldots 
\lab{lm1flows}
\ee
The following result applies to this case \ct{v502} :
\begin{lemma}
Equations \rf{lmflows} imply that
$\p_i \in {\rm Ker} \,L_m$ and $\psi_i \in {\rm Ker} \,L_m^{\ast}$
are ``up to a gauge rotation'' (adjoint) eigenfunctions satisfying:
\br 
\pa_n \p_i & = & B_n (\p_i) \qquad i= 1, \ldots ,m
\lab{panpi}\\
\pa_n \psi_i & = &- {\wti B}_n^{\ast} (\psi_i) \qquad 
i= 1, \ldots ,m
\lab{panpsii}
\er
\label{lemma:lema}
\end{lemma}
\begin{proof}
{}From $\pa_n L_m (\p_i) = 0$ and \rf{lmflows} we find that
$ B_n  (\p_i) -\pa_n \p_i \in {\rm Ker} \,L_m$.
Hence we can write
\be
B_n (\p_i) -\pa_n \p_i = - \sum_{j=1}^m \p_j\, c_{ji}^{(n)} 
(\tit ) 
\lab{l1pipj}
\ee
where $\tit = ( t_2, t_3, \ldots)$.
We now proceed in a way similar to the one used, in a slightly different
setting, in \ct{noak}.
Define $\(\D_n\)_{jk} \equiv \pa_n \d_{jk} - c_{kj}^{(n)}$ so that we
can compactly rewrite \rf{l1pipj} as
$ \(\D_n\)_{jk} \p_k = B_n (\p_j)$.
The Zakharov-Shabat equations \rf{zsflows}, ensure the zero
curvature equation $\(\sbr{\D_n}{\D_l}\)_{ik} \p_k = 0$.
Thus the ``connection'' $c_{ij}^{(n)}$ is a pure gauge and can be cast
in a form
\be
c_{ij}^{(n)} (\tit ) = (c^{-1})_{ik} (\tit )\, \pa_n \, c_{kj}(\tit )
\qquad ; \quad n \geq 2
\lab{purgaug}
\ee
Define accordingly
\be
\bp_j \equiv \p_k \, (c^{-1})_{kj} 
\lab{bp}
\ee
It is easy to verify that $\bp_j$ satisfy
\be
\pa_n \bp_j  =  \( \D_n \p \)_k (c^{-1})_{kj}=  B_n  (\bp_j)
\lab{bpeigen}
\ee
Similarly, from $\pa_n L_m^{\ast} ( \psi_i)=0 $ we arrive at 
\be
{\wti B}_n^{\ast} (\psi_i) + \pa_n \psi_i = 
\sum_{j=1}^m {\bar c}_{ij}^{(n)} (\tit )\, \psi_j
\lab{lm1psipsj}
\ee
We will  now establish a relation between coefficients $c_{ij}^{(n)}$ and
${\bar c}_{ij}^{(n)}$.
We need at this point 
 and the technical identity: 
\be
 \lb K\, , \, f D^{-1} g \rb_{-} =K (f) D^{-1} g-  f D^{-1} K^{\ast} (g)
\lab{tkppsi}
\ee
valid for a purely differential operator $K$ and arbitrary functions
$f,g$. We find from \rf{lm1flows} and the above equation that
\be
\( \pa_n L_{m}^{-1}\)_{-} = \sum_{i=1}^{m} 
\(B_{n} (\p_i )\) D^{-1} \psi_i 
- \sum_{i=1}^{m} \p_i  D^{-1} {\wti B}_{n}^{\ast} (\psi_i)
\lab{panmin}
\ee
Equations \rf{l1pipj} and \rf{lm1psipsj} agree with \rf{panmin} provided
\be 
\sum_{i,j=1}^{m} \( c_{ij}^{(n)} + {\bar c}_{ij}^{(n)}\) 
\p_i  D^{-1} \psi_j = 0 
\lab{proofa}
\ee
Define a differential operator of $m-1$ order
\be 
K \lb \p \rb \; \equiv \; 
\sum_{s=1}^m \sum_{l=0}^{s-1}  u_l D^l ( \p)^{(s-l-1)}
\lab{proofb}
\ee
such that $K^{\ast} \lb \p \rb (\psi) = {\me{\p}{\psi}}_{L_m} $.
{}From \rf{proofa} and \rf{resdij} we find
\be 
\( \sum_{i,j=1}^{m} \( c_{ij}^{(n)} + {\bar c}_{ij}^{(n)}\) 
\p_i D^{-1} \psi_j\, K \lb \p_k \rb\)_{-}  = 0 
\lab{proofc}
\ee
or
\be
\sum_{i=1}^{m} \( c_{i\,k}^{(n)} + {\bar c}_{i\,k}^{(n)}\) 
 \p_i  =0 \qquad; \;\;  k=1, \ldots , m
\lab{proofd}
\ee
Since $\{ \p_i \}$ are linearly independent
we find from \rf{proofd}
that $c_{i\,j}^{(n)}= - {\bar c}_{i\,j}^{(n)}$ for all $i,j=1,\ldots ,m$.
Accordingly,
$\(\D_n^{\ast}\)_{jk}\psi_k =- \( \cL_{m+1}^{\ast}\)^{n/r}_{+} (\psi_j)$,
with $\(\D_n^{\ast}\)_{jk} \equiv \pa_n \d_{jk} + c_{jk}^{(n)}$.
Define, next
\be
{\bar \psi}_j \equiv c_{jk} \, \psi_k 
\lab{bpsi}
\ee
It follows that 
\be
\pa_n {\bar \psi}_j  =  c_{jk} \( \D_n^{\ast} \psi \)_k =
- {\wti B}_{n}^{\ast} ({\bar \psi}_j )
\lab{bpsieigen}
\ee
Hence we succeeded to find a mutually inverse gauge rotations
taking $\p_i \in {\rm Ker} \,L_m$ and $\psi_i \in {\rm Ker} \,L_m^{\ast}$
into (adjoint) eigenfunctions satisfying 
\rf{panpi} and \rf{panpsii}
\end{proof}

\begin{lemma}
Let $\p$ and $\psi$ satisfy \rf{panpi} and \rf{panpsii} with respect to
flows from \rf{lmflows}, then
\br
\pa_n {\me{\p}{\psi}}_{L_m} &=& 
{\rm Res} \( D^{-1} \psi {\wti B}_n L_m (\p) D^{-1} \)
- {\rm Res} \( D^{-1} L_m^{\ast} (\psi)  B_n \p D^{-1} \) \nonu \\
&=& {\me{L_m (\p)}{\psi}}_{{\wti B}_n} -
{\me{\p}{L_m^{\ast} (\psi)}}_{B_n} \quad \quad n=1,2,{\ldots} 
\lab{clemma}
\er
\label{lemma:cc}
\end{lemma}
\begin{proof}
Proof follows from the technical Lemma \ct{oevel93}:
\be
{\rm Res} \( D^{-1} L_1 L_2 D^{-1} \) =
{\rm Res} \( D^{-1} \(L_1^{\ast}\)_{0} L_2 D^{-1} \) +
{\rm Res} \( D^{-1} L_1 \(L_2\)_{0} D^{-1} \) 
\lab{oevel93}
\ee
Where $L_1, L_2$ are arbitrary differential operators and $\(\cdot\)_{0}$
denotes projection on the zero-order term.
With the help of relation \rf{oevel93} we can rewrite 
${\rm Res} \( D^{-1} \psi L_m  B_n \p D^{-1} \) $ as
a sum of ${\rm Res} \( D^{-1} L_m^{\ast} (\psi)  B_n \p D^{-1} \)$
and ${\rm Res} \( D^{-1} \psi L_m  B_n (\p) D^{-1} \) $.
\end{proof}

\begin{corollary}
For   $\p \in {\rm Ker}\, L_m $ and $\psi \in {\rm Ker}\, L_m^{\ast}$
and $L_m$ satisfying eq. \rf{lmflows}
it holds that $\pa_n {\me{\p}{\psi}}_{L_m} =0$ for $n=1,2,{\ldots} $. 
Accordingly, ${\me{\cdot}{\cdot}}_{L_m} $ defines a canonical pairing 
$ {\rm Ker} \,  L_m \times {\rm Ker} \,L_m^{\ast} \to \IC$.
\label{corollary:btwo}
\end{corollary}

As a special case ($n=1$) of Lemma \ref{lemma:cc} we have
equation 
\be
\pa_x  {\me{\p}{\psi}}_{L_m} \,= \,L_m \( \p\) \, \psi - 
\p \, L_m^{\ast} \( \psi\)
\lab{bila}
\ee
Note, that the result \rf{bila} is valid this time for an arbitrary 
$\p, \psi$ as follows by verification.

Another consequence of Lemma \ref{lemma:cc} reads :
\begin{corollary}
For $\p, \psi$ satisfying condition of Lemma \ref{lemma:cc} and
$L_m$ whose isospectral flows are given in \rf{lmflows} the following
relation holds:
\be
{\me{\p}{\psi}}_{L_m} = S \({L_m (\p)},{\psi}  \) \,-\,
S \( {\p}, {L_m^{\ast} (\psi)} \)
\lab{frommetos}
\ee
up to a constant (in the multi-time $t$).
\label{corollary:bthree}
\end{corollary}
Eq.\rf{frommetos} follows from eq.\rf{clemma} and relations:
\be
\pa_n S \({L_m (\p)},{\psi}  \) = {\me{L_m (\p)}{\psi}}_{{\wti B}_n}
\quad ; \quad 
\pa_n S \( {\p}, {L_m^{\ast} (\psi)} \)= {\me{\p}{L_m^{\ast} (\psi)}}_{B_n} 
\lab{twos}
\ee

{\sect{Lax operator representation of the CKP hierarchy}}

We are studying here the class of constrained \cKP 
hierarchies for which we have the Lax representation: 
\be
L  =   D^r+ \sum_{l=0}^{r-2} u_l D^l +   \sum_{i=1}^m \P_i D^{-1} \Psi_i
= B_r +   \sum_{i=1}^m \P_i D^{-1} \Psi_i
\lab{ckprm}
\ee
with $\P_i, \Psi_i$ being (adjoint) eigenfunctions of the Lax operator
$L$ as in \rf{eigenlax}.
As shown in \ct{v502,HL,HLa} the \cKPrm hierarchy can be expressed in terms 
of two normalized differential operators $L_m$, $L_{m+r}$ of 
order $m$ and $r+m$, respectively. 
The Lax operator \rf{ckprm} of the \cKPrm hierarchy is in this representation
being rewritten as a ratio:
\be
L  \equiv L^{-1}_m L_{m+r} = \sum_{i=1}^{m} \p_i D^{-1} 
L_{m+r}^{\ast} (\psi_i) + B_r
\lab{lckprm}
\ee
The wave eigenfunction $\psi_W (t, \l)$ of \rf{ckprm} 
is an eigenfunction (as in eq.\rf{eigenlax}) which additionally  satisfies
the following spectral equation:
\be
L \psi_W (t, \l)  \equiv B_{r} (\psi_W (t, \l)) + 
\sum_{i=1}^m \P_i (t) \, S \( \psi_W (t, \l), \Psi_i (t)\)
= \l^r \psi_W (t, \l) 
\lab{grassckp}
\ee
In eq.\rf{grassckp} $S \( \psi_W (t, \l), \Psi_i (t)\)$ 
is the squared eigenfunction potential (SEP) associated
with a pair of eigenfunctions $\psi_W (t, \l)$ and $\Psi_i (t)$.

{\sect{ Universal Sato's Grassmannian construction of 
the CKP Hierarchy}} 

Let us first introduce the following basic definition:
\begin{definition}
For the wave-eigenfunction of the KP hierarchy and  
$\psi_i \in {\rm Ker} L^{\ast}_m$ we define $m$ objects:
\be
\cS_i (t, \l ) \equiv \l^r {\me{\psi_{W}}{\psi_i}}_{L_m}
\quad ;\quad
i=1, {\ldots} ,m
\lab{condition}
\ee
\label{definition:csidef}
\end{definition}

As seen from eq. \rf{bila} the $m$ objects $\cS_i (t, \l )$ defined in 
\rf{condition} satisfy :
\be
\pa_x \cS_i (t, \l ) = \l^r \psi_i  L_m \( \psi_{W} \)\quad ;\quad
i=1,\ldots ,m
\lab{paxcsi}
\ee
Note, also that the expressions \rf{condition} and \rf{psilmic} lead
to 
\be 
\cS_i (t, \l ) = \l^r \psi_i L_{m,i} \( \psi_W \)
\lab{csii}
\ee
where $L_{m,i} $ are the ordinary differential operators of order $m-1$,
whose kernels are given by
$ {\rm Ker} \( L_{m,i}  \) = 
\{ \p_1, \ldots , {\widehat \p_{i}}, \ldots, \p_m \}$.
The action of $L_{m,i} $ is defined in eq.\rf{lmic}

The following Lemma establishes a connection between \cKPrm reduction of the KP
hierarchy and the Grassmannian formulation.

\begin{lemma}
The \cKPrm reduction within the KP hierarchy is equivalent to 
the following system defined in terms of the Grassmannian:\\
\nit
1)~ Let $\{ \P_1, \ldots , \P_m \}$ be $m$ linearly independent
functions \\
\nit
2)~ Let $m$ objects $\cS_i (t, \l )$ be defined as in
Def.\ref{definition:csidef} in terms of $\psi_i$ dual to $\Phi_i$ according to
\rf{psiipi}.
Let $\cS_i (t, \l )$, furthermore, satisfy the following two conditions :
\br
\pa_x \cS_i (t, \l )  &\in& {\bf W}
\lab{condition-a}\\
\sum_{i=0}^m c_i \cS_i (t, \l ) \in {\bf W}  \;& {\rm implies}&\;
c_i = 0
\lab{condition-b}
\er
\label{lemma:main}
\end{lemma}
\begin{proof}
We start the proof with the \cKPrm system as given in \rf{ckprm} with
both $\{ \P_i \}$ and $\{ \Psi_i \}$ being linearly independent set of
functions. 
It follows from \rf{grassckp}, \rf{wgras} and
$ \pa_x S \( \psi_W (t, \l), \Psi_i (t)\)= \psi_W (t, \l) \Psi_i (t)$
that
\be
\l^r \psi^{(j)}_W (t, \l)  -
\sum_{i=1}^m \P_i^{(j)} (t) \, S \( \psi_W (t, \l), \Psi_i (t)\)
\; \in \; {\bf W}  \quad ; \quad j=0, {\ldots} , m-1
\lab{grassa}
\ee
or in the matrix notation:
\be
\l^r \left( \begin{array}{c} 
\psi_W (t, \l) \\ \psi^{(1)}_W (t, \l) \\ \vdots \\ \psi^{(m-1)}_W (t, \l)
\end{array} \right)
- \cW_{m \times m} \left( \begin{array}{c} 
S \( \psi_W (t, \l), \Psi_1 (t)\) \\ S \( \psi_W (t, \l), \Psi_2 (t)\)
 \\ \vdots \\ S \( \psi_W (t, \l), \Psi_{m} (t)\)
\end{array} \right) \in {\bf W}
\lab{mgrassa}
\ee
meaning that each element of the above combination of columns belongs to the
Grassmannian ${\bf W}$.
One finds easily from \rf{mgrassa} and
$ \pa_x S \( \psi_W (t, \l), \Psi_i (t)\)= \psi_W (t, \l) \Psi_i (t)$
that
\be
\l^r \pa_x \cW^{-1}_{m \times m}  \left( \begin{array}{c} 
\psi_W (t, \l) \\ \psi^{(1)} (t, \l) \\ \vdots \\ \psi^{(m-1)} (t, \l)
\end{array} \right)
\in {\bf W}
\lab{magras}
\ee

In terms of the matrix elements $ \(\cW^{-1}_{m \times m}\)_{ij} $ 
from \rf{deltas} the relation \rf{magras} takes a form
\be
\l^r \pa_x \sum_{j=0}^{m-1} \(\cW^{-1}_{m \times m}\)_{ij} 
\psi^{(j)} (t, \l) \; \in \; {\bf W}
\lab{condition1}
\ee
for each $i=1,{\ldots} ,m$. This yields condition \rf{condition-a} 
due to the Lemma \ref{lemma:psiiclm}.

Recalling Lemma \ref{lemma:psiiclm} we see that \rf{mgrassa} implies
\be
S \( \psi_W (t, \l), \Psi_i   \) - \cS_i (t, \l) \in {\bf W} \quad;\quad
i=1,{\ldots} , m
\lab{smcs}
\ee
If now it holds that
\be
\sum_{i=1}^m  c_i \cS_i (t, \l) \in {\bf W}
\lab{lindepcs}
\ee
then because of relation \rf{smcs} eq.\rf{lindepcs} implies that
\be
\sum_{i=1}^m  c_i S \(  \psi_W (t, \l), \Psi_i  \) \in {\bf W}
\lab{hiSpsi} 
\ee
and therefore due to eq.\rf{S-osci} 
$\sum_{i=1}^m  c_i \Psi_i \l^{-1} \exp \xi (t, \l) =0$.
We note that from the linear independence of $\{ \Psi_1, \ldots , \Psi_m \}$
it follows that $ c_i =0$, which is the desired result.

{}From now on, we will assume that given is the system of the linearly
independent functions $ \{ \P_1, \ldots , \P_{m} \}$ together with
conditions \rf{condition-a}-\rf{condition-b}.
We are going to show that the KP hierarchy associated
with the wave-eigenfunction $\psi_W (t, \l)$ satisfying constraints
\rf{condition-a} and \rf{condition-b} belongs to the \cKPrm class.

We start by defining $m$ functions:
\be
\Psi_i (t,t_0) \equiv  \int d \l\, \psi^{\ast}_{W} (t, \l )  \, \cS_i (t_0, \l)
\quad ; \quad i=1,{\ldots} , m
\lab{def-psia}
\ee
with $\cS_i (t_0, \l)$ defined as in \rf{condition}.
First, it follows clearly from the definition \rf{def-psia}
that $\Psi_i$ is an adjoint eigenfunction in the multi-time $t$.
Secondly, $\Psi_i$
is non-zero only for $\cS_i (t_0, \l)$ not in ${\bf W}$ due to the 
Hirota's identity. According to the condition \rf{condition-b} the $m$
functions $\Psi_i$ are linearly independent.

What remains to be proven in order to establish that $\Psi_i$ from
eq.\rf{def-psia} are adjoint eigenfunctions is that the functions
$\Psi_i$ do not depend on the second multi-parameter $t_0$.
Indeed, from the condition \rf{condition-a} it follows immediately that
$\pa_{x_0} \Psi_i (t,t_0) = 0$ and accordingly $\Psi_i $ does not depend on
$x_0= (t_0)_1$.
To complete the proof it remains to show that indeed
$\pa \Psi_i / \pa (t_0)_n = 0 $ for $n>1$.

Define an oscillatory function $\psi_V (t, \l)$ by
\be
L_m ( \psi_W (t, \l) ) \equiv \l^m \psi_V (t, \l) \,.
\lab{psiv}
\ee
Since $\psi_W (t, \l) = W \exp \xi (t, \l)$ where $W$ is a dressing
operator, we find that $\psi_V (t, \l) = V \exp \xi (t, \l)$
where $V = L_m W D^{-m}$ has like $W$ a form of the dressing operator
$V=1 +\sum_{i=1}^{\infty} v_i D^{-i}$.
Alternatively, we can rewrite $L_m  = V D^{m} W^{-1}$.
Consider, $L_{m+r} \equiv V D^{m+r} W^{-1}$ such that
$L_{m+r} ( \psi_W (t, \l) ) =\l^{r}  L_m ( \psi_W (t, \l) )$.
Since $\l^{r}  L_m ( \psi_W (t, \l) ) \in {\bf W}$ the operator $L_{m+r}$
is an ordinary differential operator.
Moreover, we find that the KP Lax operator $L = W D^r W^{-1}$ can be written
in terms of two ordinary differential operators as
$L = L_m^{-1} L_{m+r}$.
{}From \ct{krichev} and \ct{dickey95} we know
that the KP hierarchy equations $\pa_n L = \sbr{\( L \)^{n/r}_{+}}{L}$
for $L = L_m^{-1} L_{m+r}$ are equivalent to the following flows on the
differential operators $L_m, L_{m+r}$ :
\br
\pa_n L_m &=& \( L_{m+r} L_m^{-1} \)^{n/r}_{+} L_m - 
L_m \( L_m^{-1} L_{m+r} \)^{n/r}_{+} \lab{lmflo}\\
\pa_n L_{m+r} &=& \( L_{m+r} L_m^{-1} \)^{n/r}_{+} L_{m+r} - 
L_{m+r} \( L_m^{-1} L_{m+r} \)^{n/r}_{+} \lab{lmrflo}
\er
It has been shown in \ct{v502} that equations \rf{lmflo} and \rf{lmrflo} 
imply that $\p_i \in {\rm Ker} \,L_m$ and $\psi_i \in {\rm Ker} \,L_m^{\ast}$
are ``up to a gauge rotation'' (adjoint) eigenfunctions satisfying:
\br 
\pa_n \p_i & = &\( L_m^{-1} L_{m+r} \)^{n/r}_{+}(\p_i) 
\qquad i= 1, \ldots ,m
\lab{panpil}\\
\pa_n \psi_i & = &- \( (L_{m+r} L_m^{-1})^{\ast}  \)^{n/r}_{+} (\psi_i) \qquad 
i= 1, \ldots ,m
\lab{panpsiil}
\er
We recognize in the above equations the setting of Lemma \ref{lemma:lema}
with \rf{panpil}-\rf{panpsiil} appearing to be special cases of 
\rf{panpi}-\rf{panpsii}.
Especially, we may use the results of Lemma \ref{lemma:cc} and equation
\rf{clemma} to find
\be
\pa_n \cS_i = \l^r
{\me{L_m (\psi_W)}{\psi_i}}_{\( L_{m+r} L_m^{-1}  \)^{n/r}_{+}}
= \l^r A_{n-1} \(L_m (\psi_W)\)
\lab{anm1}
\ee
with $A_{n-1}$ being $(n-1)$-th order differential operator.
Since $ \l^r A_{n-1} \(L_m (\psi_W)\) \in {\bf W}$ it follows
immediately that $\pa \Psi_i / \pa (t_0)_n =0$ and $\Psi$ is indeed 
a function of the multi-time $t$ only.

Note, that on basis of relation \rf{psiiclm} the alternative form of the 
definition \rf{def-psia} appears to be:
\be
\Psi_i (t) \equiv  \sum_{j=0}^{m-1} \(\cW^{-1}_{m \times m}\)_{ij} (t_0) 
\int d \l \l^r \psi^{\ast}_W (t, \l )  \, \psi^{(j)}_{W} (t_0, \l)
\lab{def-psiaa}
\ee
Accordingly, using \rf{deltas} we find (see also \ct{cheng})
\be 
\sum_{i=1}^{m} \P_i (t_0) \Psi_i (t) = 
\int d \l\, \l^r \psi^{\ast}_W  (t, \l )  \, \psi_W (t_0, \l)
\lab{ppsi}
\ee
from which it follows that
\be
\sum_{i=1}^{m} \( \pa_n \P_i (t_0)- B_n \(\P_i (t_0) \) \) \Psi_i (t) = 0
\quad ; \quad n=1,2, {\ldots} 
\lab{lin-ind}
\ee
or equivalently
\be
\sum_{i=1}^{m} \( \pa_n \P_i (t_0)- B_n \(\P_i (t_0) \) \) \cS_i (t, \l) \in
{\bf W}
\lab{lin-inde}
\ee
{}From the last identity \rf{lin-inde} and condition 
\rf{condition-b} we conclude that $\P_i$ are 
eigenfunctions for $i= 1,{\ldots}, m$.

Recall, that
$ L \psi_W (t, \l) = \l^r \psi_W (t, \l) = L_{+} (\psi_W (t, \l))+ 
L_{-}( \psi_W (t, \l))$ 
with the pseudo-differential part
$ L_{-} (\psi_W (t, \l)) \sim O \( \l^{-1} \) \exp \xi (t, \l ) $.
Inserting it back into eq.\rf{ppsi} we find
\be 
\sum_{i=1}^{m} \P_i (t_0) \Psi_i (t) = 
\int d \l \psi^{\ast}_W  (t, \l )\,L_{-} (\psi_W (t_0, \l))
\lab{ppsia}
\ee
{}From \ct{sep}  we conclude that \rf{ppsia} implies
\be
L_{-} (\psi_W (t_0, \l))=
\sum_{i=1}^{m} \P_i (t_0)  
{1 \o \l } \psi_W (t_0, \l) \Psi_i \(t_0 - \lb \l^{-1} \rb \)
\lab{lmin}
\ee
up to terms in ${\bf W}$.
Equivalently, we can rewrite the last relation in the desired form
\be
L_{-} (\psi_W (t, \l))=
\sum_{i=1}^m  \P_i (t) S \( \psi_W (t, \l), \Psi_i (t )\)
\lab{lmina}
\ee
from which eq.\rf{ckprm} follows due to the fact that the pseudodifferential
operators act freely on the wavefunctions as seen from \rf{dinverse}.
\end{proof}

{\sect{Truncated KP Hierarchy as cKP Hierarchy}}

Let us consider the truncated KP hierarchy defined by the dressing operator
$W$ containing only finite number of terms.
Let $K$ be a positive order differential operator of the order
$N$, such that $N>m$ and such that $W= KD^{-N}$.
Accordingly, the corresponding Lax operator is
\be
L_{\rm tr} = K D^r K^{-1} = W D^r W^{-1}
\lab{trun}
\ee
Let $f_i, g_i$ with $i=1,{\ldots} ,N$ be elements of the kernels
${\rm Ker} K$ and ${\rm Ker} K^{\ast}$, respectively.
As shown in \ct{noak} the Wilson-Sato equations \rf{sato-a} for the
hierarchy defined by the Lax operator \rf{trun} take a simple form
for the elements of  ${\rm Ker} K$:
\be
\pa_n f_i = \pa_x^n f \quad ;\quad i =1, {\ldots} , N
\lab{nfi}
\ee

We have 
\be
K^{-1} = \sum_{i=1}^N f_i D^{-1} g_i
\lab{kinverse}
\ee
and consequently
\br
{\rm Res} \(K D^r K^{-1}\) &= &\sum_{i=1}^N K D^r(f_i) g_i =
\sum_{i=1}^N (-1)^{N-i}  { W \lb f_1, \ldots , f_N, f^{(r)}_i  \rb 
  W \lb f_1, \ldots , {\widehat f_{i}}, \ldots 
f_N  \rb \o W^2 \lb f_1, \ldots , f_{N}\rb } \nonu\\
&= & \pa_x \sum_{i=1}^N (-1)^{N-i} 
{   W \lb f_1, \ldots , {\widehat f_{i}}, \ldots 
f_N, f^{(r)}_i  \rb \o W \lb f_1, \ldots , f_{N}\rb }
= \pa_x \pa_r \ln W \lb f_1, \ldots , f_{N}\rb \lab{trutau}
\er
where we used the Jacobi identity 
\be
W \Blb W \lb f_1, {\ldots} , f_m, g \rb , W \lb f_1, {\ldots} , f_m, h \rb
\Brb = W \lb f_1, {\ldots} , f_m \rb W \lb f_1, {\ldots} , f_m, g,h \rb
\lab{jac}
\ee
and \rf{nfi}.

Hence we reproduced the well-known result that the tau function for 
the truncated KP hierarchy is the Wronskian
$\t_{trun} = W \lb f_1, \ldots , f_{N}\rb$.

Due to \rf{nfi} we can rewrite $ f_i$ as: 
$ f_i = \int dz {\ti f}_i (z) \exp \( \xi (t, z)\)$.
Notice, that
\be
K \exp \( \xi (t, \l)\) = z^{N} \psi_W (t, \l)
\lab{kdress}
\ee
due to the fact that $W= KD^{-N}$ is the dressing operator of the truncated 
hierarchy.
It holds therefore that 
\be
 0= K (f_i) = \int dz z^N {\ti f}_i (z) \psi_W (t, \l)
\lab{kfi}
\ee
Accordingly, for any positive differential operator $B$ we find
\be 
\int dz z^N {\ti f}_i (z) B \(\psi_W (t, z) \) = 
B \( \int dz z^N {\ti f}_i (z) \psi_W (t, z) \) = 0
\lab{kfia}
\ee

We now investigate the condition for $L_{\rm tr} $ to be
within \cKPrm in the nontrivial case $N>m$.
As shown above, the necessary condition for this to happen is that
$\l^r L_m (\psi_W) \in {\bf W}$ or that there exists a
positive differential operator $B$ such that
$\l^r L_m (\psi_W) = B (\psi_W)$.
Comparing with \rf{kfia} we find that $ \l^r L_m ( \psi_W ) \in {\bf W}$
translates into:
\br
0&=& \int dz z^N {\ti f}_i (z) z^r L_m \( \psi_W (t, z)\) =
L_m \(\int dz {\ti f}_i (z) z^{r+N} \psi_W (t, z)\) \nonu\\
&=& L_m \(\int dz {\ti f}_i (z) z^{r}K  e^{\xi (t, z)}\)=
L_m K D^r (f_i)
\lab{lmkrfii}
\er
for all $i=1, {\ldots} , N$.

Hence $ K D^r (f_i) \in {\rm Ker} (L_m)$ for $i=1,{\ldots} , N$.
In \ct{oevel96} this condition was rewritten using a Jacobi identity for
Wronskians as
\be
W \lb f_1, \ldots , f_{N}, f_{i_1}^{(r)} , {\ldots}, f_{i_{m+1}}^{(r)}\rb =0
\lab{oev}
\ee

\sect{Concluding Remarks}

We have seen that formulating the constrained KP hierarchy  within
the Sato Grassmannian becomes transparent when use is being made 
of the underlying ordinary differential operators with convenient
parametrization of their kernels.
Useful insight has been obtained by relating the notions of kernel elements
of the underlying ordinary differential operators with that of the 
eigenfunctions of the KP hierachy.
The related connection of the bilinear concomitant introducing the canonical 
pairing structure on the space kernels to that of the squared eigenfunction 
potentials has then arisen naturally.

Let us complete our discussion by the following additional comments
addressing fundamental questions of the formalism.

\mark Due to relation \rf{psiiclm} we have
$\sum_{i=1}^m \P_i \cS_i = \l^r \psi_W$.
Hence condition \rf{condition-b} can be understood as an obstruction to the
usual KdV reduction with $\l^r \psi_W \in {\bf W}$.

\mark Alternatively to \rf{condition-b} we could have expressed 
the relevant assumption in terms of the integrals :
\be 
\sum_{i=0}^m c_i \int d \l\, \psi^{\ast}_{W} (t, \l )  \,  \cS_i (t_0, \l) =0
\;\, {\rm implies}\,\; c_i = 0
\lab{condition-c}
\ee
instead of involving the Sato Grassmannian ${\bf W}$ in \rf{condition-b}.
The arguments used in the proof would then have worked with small adjustments
but without any need of making an additional assumption that
$ \int  d\l \,\psi_{W}^{*}(t,\l)  F \(\psi_W (t^{\pr} ,\l)\) =0$
implies $F \(\psi_W (t^{\pr} ,\l)\)\in {\bf W}$ for
$F \(\psi_W (t^{\pr} ,\l)\)$ of the positive order.

\mark {}From the definition \rf{psiv} and \rf{lmflo} we find that
the flows of $\psi_V (t, \l)$ are given by:
\be
\pa_n \psi_V (t, \l)  =  \( L_{m+r} L_m^{-1} \)^{n/r}_{+} (\psi_V (t, \l)) 
\lab{psivtn}
\ee
Based on \rf{panpsii} and \rf{psivtn} it makes now sense to define the squared
eigenfunction potential $S (\psi_V, \psi_i)$ for $\psi_V (t, \l)$ and $\psi_i$
with the following useful property:
\be
\cS_i (t, \l) = \l^{r+m} S (\psi_V, \psi_i) 
\lab{sepforcsi}
\ee

Due to eq. \rf{sepforcsi} one can rewrite eq.\rf{def-psia} as 
\be
\Psi_i (t) =  \int d \l\, \l^{r+m} \psi^{\ast}_{W} (t, \l )  \, 
S (\psi_V (t_0, \l) , \psi_i(t_0))  
\quad ; \quad i=1,{\ldots} , m
\lab{psias}
\ee
Since $\pa_{x_0} \Psi_i (t,t_0) = $ it holds that
$\int d \l\, \l^{r+m} \, \psi^{\ast}_{W} \, \psi_V\, = 0$.

Furthermore it is easy to see that
$\l^{r+m} \psi^{\ast}_{W} (t, \l )= L_{m+r}^{\ast} \( \psi^{\ast}_{V} (t, \l
) \)$ as follows from the definition \rf{BA-adjoint} and
$\psi^{\ast}_{V} (t, \l ) = V^{*\,-1} \exp {-\xi (t,\l )}$
together with $L_{m+r}^{\ast} = W^{*\,-1} (-D)^{m+r} V^{*}$.
Plugging it back in \rf{psias} we obtain :
\be
\Psi_i (t) =  \int d \l\, L_{m+r}^{\ast}\( \psi^{\ast}_{V} (t, \l ) \) \, 
S (\psi_V (t_0, \l) , \psi_i(t_0))  = L_{m+r}^{\ast} \( \psi_i \)
\lab{psiasa}
\ee

\mark
The inclusion $\pa_x \cS_i \in {\bf W}$
can be rewritten as
$ \l^r L_m ( \psi_W ) \in {\bf W}$, where $L_m$ is a $m$-order
differential operator whose action on the wave function $\psi_W $
can be viewed as $m$ successive \DB transformations.
With the kernel of $L_m$ being $\{ \Phi_1, ..., \Phi_m \}$, let
$w_j$ be such that
$ \Phi_j = \int d\l \l^{-1} (\psi_W  w_j) $ for $j=1, ...,m$.
Accordingly, $w_j$
are orthogonal to the subspace
${\bf W^{\pr}} \equiv {\rm span} \{ L_m ( \psi_W )\}$ of ${\bf W}$
with respect to the
inner product $\langle u | v \rangle \equiv \int d\l \l^{-1} u v$.
Hence,
the inclusion $ \l^r {\bf W^{\pr}} \in {\bf W}$ is a co-dimension 
$m$ inclusion.
Also, the condition \rf{condition-b} in view of \rf{csii}
expresses the assumption about the codimension $m$ being optimal.
This establishes link to the formalism of \ct{HL,HLa}.

\lskip
{\bf Acknowledgements}$\;$ I am grateful to L. Dickey and J. van de Leur for
reading the early versions of the manuscript and helpful comments.
\mskp

\end{document}